# Super-oscillating Electron Wave Functions with Sub-diffraction Spots

Roei Remez[*†1], Yuval Tsur[†1], Peng-Han Lu[2], Amir H. Tavabi[2], Rafal E. Dunin-Borkowski[2] and Ady Arie[1]

[1] School of Electrical Engineering, Fleischman Faculty of Engineering, Tel-Aviv University, Israel
[2] Ernst Ruska-Centre for Microscopy and Spectroscopy with Electrons and Peter Grünberg Institute, Forschungszentrum Jülich, Jülich, Germany

[†]These authors contributed equally to this work
*Corresponding author: roei.remez@gmail.com

Almost one and a half centuries ago, Ernst Abbe [1] and shortly after Lord Rayleigh [2] showed that when an optical lens is illuminated by a plane wave, a diffraction-limited spot with a radius $0.61\lambda/sin\alpha$ is obtained, where $\lambda$ is the wavelength and $\alpha$ is the semi-angle of the beam's convergence cone. However, spots with much smaller features can be obtained at the focal plane when the lens is illuminated by an appropriately structured beam. Whereas this concept is known for light beam, here, we show how to realize it for massive-particle wave function of a free electron. We experimentally demonstrate an electron central spot of radius 106 pm, which is more than two times smaller than the diffraction limit of the experimental setup used. In addition, we demonstrate that this central spot can be structured by adding orbital angular momentum to it. The resulting super-oscillating vortex beam has a smaller dark core with respect to the regular vortex beam. This new family of electron beams having hot-spots with arbitrarily small features and tailored structure can be useful for studying electron-matter interactions with sub-atomic resolution.

Among the methods that are used in light optics for circumventing the diffraction limit are near field microscopy [3], metamaterial-based perfect lenses and super-lenses [4] and various other super resolution schemes [e.g., 5,6]. However, none of these methods have been demonstrated with matter (*e.g.*, electron) waves. An interesting proposal for manifesting *arbitrarily* small spots for optical microscopy was made in 1952 by Toraldo di Francia [7]. Following earlier work in the microwave regime [8], he proposed putting a series of concentric rings near the lens pupil, thereby modulating the incoming wave so that the central focal spot could be made smaller than the Abbe-Rayleigh limit, accompanied by a peripheral ring of light. In a related development and following concepts that were developed for weak measurements in quantum systems [9], Michael Berry introduced the concept of super-oscillating functions and predicted their potential applications for super-resolution microscopy [10]. These super-oscillating functions are band limited functions that locally oscillate faster than their highest Fourier component [9, 10]. They have been applied successfully in light microscopy [11] for enhancing barely resolved objects, as well as for other applications involving free-space optical beams [12], nonlinear frequency conversion [13] and surface plasmon polaritons [14]. Super-oscillations have also been studied in the time domain for applications such as time-dependent sub-diffraction focusing [15] and "super-transmission" through absorbing media [16]. The concept of super-oscillating waves was not applied till now to matter waves, but it can offer attractive opportunities owing to the much shorter wavelengths of these waves with respect to optical waves. In this Letter we concentrate on electron waves and address the following questions: How to generate a super-oscillating wave function? What is the size that can be reached with comparison to the diffraction limited spot size? Can we obtain hot-spots that are comparable with the size of the atom? What are the limitations on the minimum feature size? And finally, can we structure these super-oscillating beams, for example by adding orbital angular momentum to them, thereby generating super-oscillating vortex beams?

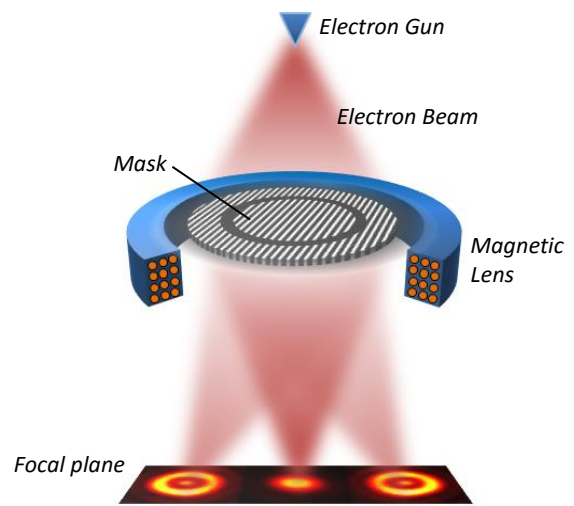

**Figure 1.** Schematic description of super-oscillating electron wave function generation. The desired wave function is created in the +1 and -1 diffraction orders.

Since the beginning of the present decade, the use of holographic masks in electron microscopy has attracted increasing attention, as it allows complete control over the electron amplitude and phase distributions, thus realizing special electron beams [17–19], and opening new possibilities for structured illumination electron microscopy [20,21]. Here, we apply such a mask to form a super-oscillatory electron wave function for the first time, and discuss its prospects. This wave function, which is designed following a simple analytic derivation [22], features a central spot that can theoretically be made arbitrarily small and routinely smaller than the Abbe-Rayleigh diffraction limit. We note that electron microscopy provides significantly higher resolution with respect to optical microscopy, since the de Broglie wavelength of the electron that we use is only 2 pm (300 keV), more than 5 orders of magnitude smaller than the wavelength of visible light. The resolution of electron microscopes is determined by a tradeoff between lens aberrations (in particular spherical aberration, which is proportional to $\alpha^3$) and diffraction (which is proportional to $1/\alpha$) [23].

We now derive a holographic mask design for producing a super-oscillatory spot. Unlike a conventional probe-forming lens, which utilizes a uniform, circular hard aperture of diameter $D$, our proposed super-oscillatory probe utilizes a transparency-phase mask $\psi_{mask}(r)$, where r is the radial distance from the column axis. We use a function previously used in optics [22,24]:

$$\psi_{mask}(r) \propto \begin{cases} e^{i\pi} & r \leq r_\pi \\ 1 & r_\pi < r \leq r_{max} \end{cases}, \quad (1)$$

where $0 < r_\pi < r_{max}$ and $r_{max} = D/2$. The transmitted wave, i.e. $\psi_{mask}(r)$, is then condensed by a lens to form the probe wave function, which is proportional to $FT\{\psi_{mask}(r)\}_\theta$ [17,25], where the angular coordinate $\theta$ is replaced (under the small-angle approximation) by $\theta = r/(f\lambda)$:

$$\psi_{probe}(a) \propto a^{-1}\left(r_{max}J_1(r_{max}a) - 2r_\pi J_1(r_\pi a)\right), \quad (2)$$

where $a = 2\pi r/f\lambda$ and $f$ is the condenser lens focal length. The probability density for electron detection at a normalized distance $a$ from the axis is given by $|\Psi(a)|^2$.

We have implemented this probe-forming mask as an off-axis computer-generated hologram [26]. Among the advantages of this method are the realization of both the phase and the amplitude of the wave function by a binary pass-block mask. The off-axis carrier wave-number $k_c = 2\pi/500 nm$ throughout this work, determines the spatial separation between the unwanted zero order and the target super-oscillatory first order, in the following computer-generated hologram expression

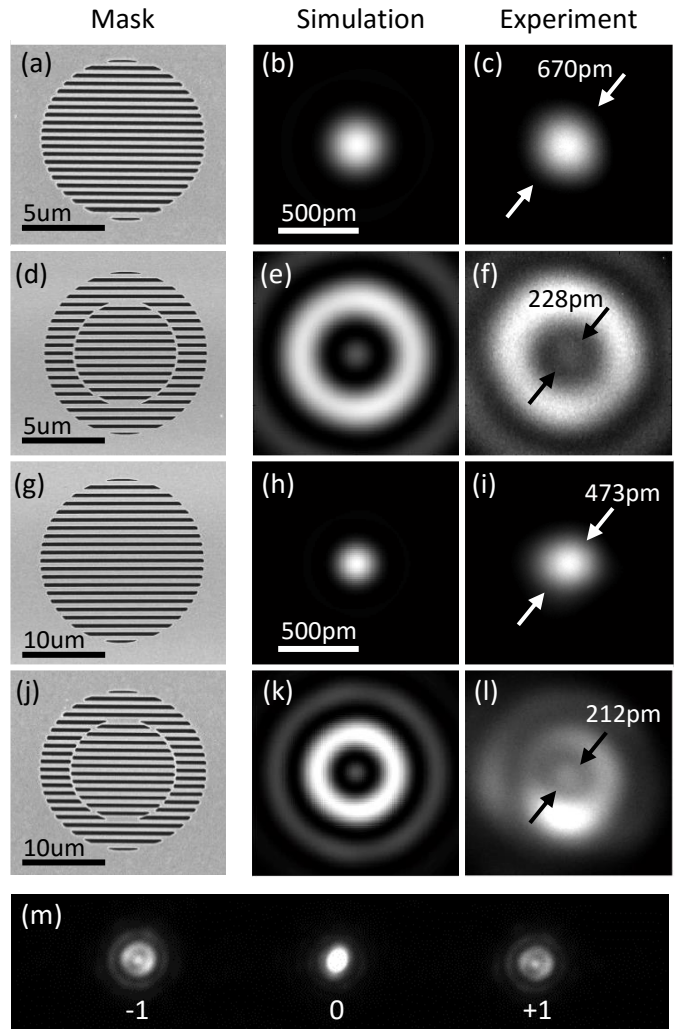

**Figure 2.** SEM images of binary amplitude masks (a,d,g,j), shown alongside simulations (b,e,h,k) and experimental results (c,f,i,l) for a circular aperture and super-oscillation masks of diameter 10 μm for $\alpha = 3.7$ mrad (rows 1 and 2), and for diameter 20 μm and $\alpha = 5.2$ mrad (rows 3 and 4). The measurements of super-oscillating electron beams (f) and (l) exhibit central hot-spots of radius 114 and 106 pm, respectively, compared to diffraction limit Airy disk radii of 334 and 235 pm, respectively. (m) – Three central orders of the diffraction pattern of the mask in (d).

$$\psi_{holo}(r,y) \propto \begin{cases} sign(\cos(k_c y + \pi)) + 1 & r \le r_\pi \\ sign(\cos(k_c y)) + 1 & r_\pi < r \le r_{max} \end{cases} \quad (3)$$

As illustrated in Fig. 1, the electron wave function will be the Fourier transform of $\psi_{holo}(r,y)$, for which orders +1 and -1, have the form of eq. (2):

$$\Psi_{probe}^{\pm}(a_\pm) \propto a_\pm^{-1}\left(r_{max} J_1(a_\pm r_{max}) - 2r_\pi J_1(a_\pm r_1)\right), \quad (4)$$

where $\vec{r} = (x,y)$, $x, y$ are the Fourier plane coordinates, $\vec{r}_0 = \left(0, \frac{k_c f}{k}\right)$, and $a_\pm \equiv \frac{k|\vec{r} \pm \vec{r}_0|}{f}$.

We note that it is also possible to realize the desired pattern on the column axis ($r = 0$) using a variable thickness phase mask [21]. In addition, although the outer ring of the super-oscillatory spot resembles a vortex beam [27], there is no helical phase present, as evident from Eq. (2), which shows that the probe is purely real. Figs. 2f and 2l present experimentally realized super-oscillatory electron probes, which have central hot-spot radii of 114 pm and 106 pm, i.e., 66% and 55% smaller than the diffraction limit Airy radii of 334 pm and 235 pm for the convergence semi-angles used $\alpha = 3.7$ mrad and 5.2 mrad, respectively. Note that this hot spot is 3 orders of magnitude smaller with respect to those that were obtained in light optics [11] and is comparable to the size of small atoms [28,29]. The experimental details can be found in the Supplemental Material.

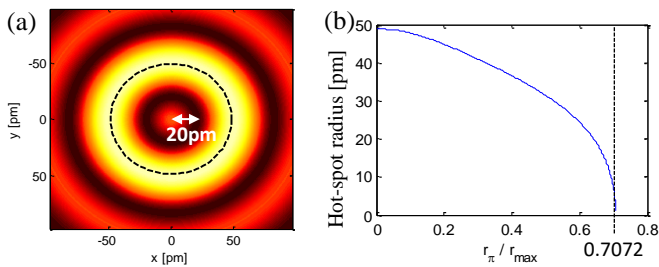

**Figure. 3.** (a) Simulation of a super-oscillating electron beam for an aberration corrected microscope in both the probe and the imaging parts with $\alpha = 25$ mrad and $r_\pi / r_{max} = 0.64$. The radius of the central hot-spot, 20 pm, is only 10 times the electron wavelength. The dashed black line marks the diffraction limit Airy disk. (b) Central lobe radius plotted as a function of $r_\pi / r_{max}$. Theoretically, an arbitrarily small electron hot-spot can be achieved, as long as factors such as spatial coherence and signal to noise ratio are addressed.

Our experiment was limited by the imaging objective lens of the confocal setup, whose spherical aberration could not be corrected. The spherical aberration of this uncorrected lens explains why, for the larger convergence angle experiment (Fig. 2(g-l)), the circular aperture spot was significantly larger than the theoretical diffraction limited spot (comparing Fig. 2i with Fig. 2h). In the Supplemental Materials Section, we show a similar experiment having a much smaller convergence angle, hence with negligible spherical and chromatic aberrations, as well as a negligible effect of partial spatial coherence and inelastic scattering. In that case, the experimental spot size is in excellent agreement with the theoretical prediction, although the smaller convergence angle also leads to a much larger overall spot size.

Let us now consider the fundamental limitations for the hot-spot size. In Fig. 3, we calculate the size of the super-oscillating central hot-spot assuming that both the probe-forming and the imaging parts of the microscope are aberration-corrected for a convergence semi-angle of 25 mrad, slightly more conservative than the value of $\alpha = 32.4 \ mrad$ achieved by Sawada *et. al.* [30]. In addition, we consider here also incoherent aberrations, namely partial spatial coherence and chromatic aberration. In the Supplemental Materials Section, we show that both incoherent aberrations are small enough using present technology to create the probe suggested in Fig. 3a, with a radius of 20 pm, only 10 times the electron wavelength. In contrast to conventional spots, which are bounded by diffraction associated with the aforementioned Airy disk, the super-oscillating spot can be arbitrary small for large enough values of the phase jump radius $r_\pi$ (see Fig. 3b). This conceptually includes hot-spots that are smaller than the electron wavelength, as was already demonstrated in light optics [11], although significant limitations stem from the low relative intensity of the hot-spot, as shown in Fig. S4 in the Supplementary Material Section, as well as from the mechanical stability of the microscope column.

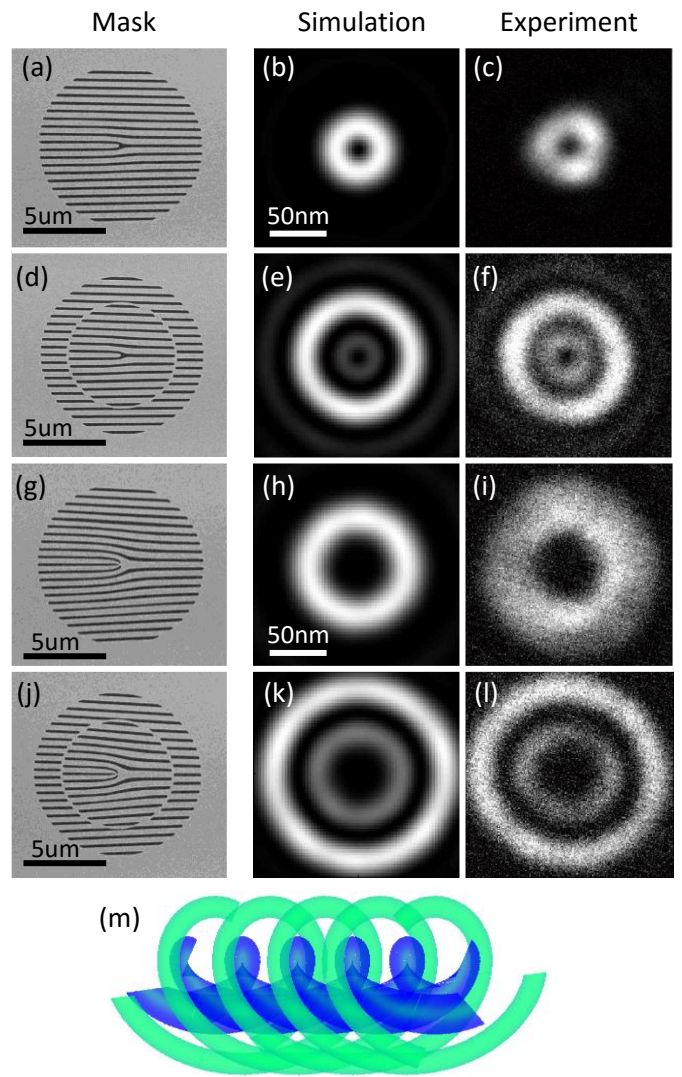

**Figure 4.** Super-Oscillating electron vortex beams. Vortex and super-oscillating vortex beams with OAM=1 (rows 1 and 2, respectively) and OAM=3 (rows 3 and 4). (m) – Schematic description of equi-phase surfaces of OAM=3 super-oscillatory electron vortex (row 4), with inner and outer rings colored with blue and green, respectively.

As demonstrated recently by Singh et. al [31] in light optics, the central spot of the super-oscillating beam can be structured. Here we implement this concept in electron optics to create super-oscillating electron vortex beams. Such vortex beams are characterized by a helical phase and carry Orbital Angular Momentum (OAM). Owing to the on-axis phase singularity, these beams are characterized by a doughnut shape, with a dark central core. The generation of electron vortex beams [17,21,25,27] and their application for studying magnetic dichroism [17,32] and for rotating nano-particles [33] have generated much interest in recent years. The super-oscillating vortex beams as shown here have smaller dark cores compared to conventional vortex beams, thereby potentially enabling to study the transfer of angular momentum between electrons and matter with improved resolution.

To generate these beams, we used the following mask design:

$$\psi_{holo}(r, y) \propto \begin{cases} sign(\cos(k_c y + l\phi + \pi)) + 1 & r \leq r_\pi \\ sign(\cos(k_c y + l\phi)) + 1 & r_\pi < r \leq r_{max} \end{cases}$$
(5)

Where $\phi$ is the angle with y axis at the plane of the mask and l is an integer. In comparison to eq. 1 that was used to generate the non-structured super-oscillating beam, the helical phase term $\exp(il\phi)$ was added. This adds an OAM of $\hbar l$ per electron. In Fig. 4 (f) and (i), we show in experiment an electron beams with OAM of $1\hbar$ and $3\hbar$, having inner rings which are respectively smaller by 20% and 32% compared to conventional beams with same OAM values. This experiment was done using $\alpha = 0.05 mrad$ and mask diameters of 10 micron, but could easily be repeated for larger convergence angle similarly to Fig. 2. As shown by Singh et al [31] for the case of light beams, other structures, with interesting properties, can be imposed on the super-oscillating hot spot – for example Airy functions that exhibit self-acceleration, or multi-lobed Hermite-Gauss functions.

In this Letter, we have presented for the first time super-oscillating wave functions for a massive particle, featuring a central spot that is smaller than the Abbe-Rayleigh diffraction limit, as well as super oscillating vortex beams. These can be the first members in a new family of shaped super-oscillatory electron wave-functions. The transmission electron microscope operates fundamentally in the single-particle regime, supporting the assertion that super-oscillation stems from interference of the wave function with itself [34]. We have shown a straightforward, highly efficient method to produce the wave function, generated for an arbitrary focal length. We have demonstrated how a hot-spot only 10 times the electron wavelength can be created nowadays using this method. In addition, with improving technology, a sub-wavelength electron hot-spot might become possible. We believe that this demonstration opens a wide range of possibilities in electron wave function manipulation, in order to create dense oscillations that were previously thought impossible in finite electron wave functions, such as sub-diffraction needles [35] that can enable increased electron beam lithography resolution, super-oscillatory shape-preserving beams [36], and enhanced electron-microscope images as was demonstrated with light [11,37,38]. Arbitrarily dense electron wave functions, having a tailored shape, can be created by our method using prolate spherical wave functions [39] (with a penalty to the average amplitude in the super-oscillating area). The super-oscillation wave function can also be used as an initial state for the realization of weak measurements of displacement in quantum systems [38, 39]. In addition, we note that super-oscillations could be created using the concept presented here with other massive particles and even large molecules [42].

The authors would like to acknowledge Prof. Yakir Aharonov, Prof. Amit Kohn and Prof. Hannes Lichte for critical discussions. This work was supported by DIP, the German-Israeli Project cooperation and by the Israel Science Foundation, grant no. 1310/13. The research leading to these results received funding from the European Research Council under the European Union's Seventh Framework Programme (FP7/2007-2013) / ERC grant agreement no. 320832.

**References**

1. E. Abbe, "Beiträge zur Theorie des Mikroskops und der mikroskopischen Wahrnehmung," Arch. für Mikroskopische Anat. **9**, 413–418 (1873).

2. F. R. S. Rayleigh, "XXXI. Investigations in optics, with special reference to the spectroscope," Philos. Mag. Ser. 5 **8**, (1879).

3. U. Dürig, D. W. Pohl, and F. Rohner, "Near-field optical-scanning microscopy," J. Appl. Phys. **59**, 3318–3327 (1986).
4. J. B. Pendry, "Negative refraction makes a perfect lens," Phys. Rev. Lett. **85**, 3966–3969 (2000).
5. S. W. Hell, "Microscopy and its focal switch.," Nat. Methods **6**, 24–32 (2009).
6. W. Lukosz, "Optical Systems with Resolving Powers Exceeding the Classical Limit," J. Opt. Soc. Am. **56**, 1463 (1966).
7. G. Di Francia, "Super-gain antennas and optical resolving power," Nuovo Cim. **9**, 426 (1952).
8. S. A. Schelkunoff, "A Mathematical Theory of Linear Arrays," Bell Syst. Tech. J. **22**, 80–107 (1943).
9. Y. Aharonov, J. Anandan, S. Popescu, and L. Vaidman, "Superpositions of time evolutions of a quantum system and a quantum time-translation machine," Phys. Rev. Lett. **64**, 2965–2968 (1990).
10. M. Berry, "Faster than Fourier," in *Quantum Coherence and Reality, Celebration of the 60th Birthday of Yakir Aharonov*, J. S. Anandan and J. L. Safko, eds. (World Scientific, 1994), pp. 55–65.
11. E. T. F. Rogers, J. Lindberg, T. Roy, S. Savo, J. E. Chad, M. R. Dennis, and N. I. Zheludev, "A super-oscillatory lens optical microscope for subwavelength imaging.," Nat. Mater. **11**, 432–5 (2012).
12. E. Greenfield, R. Schley, I. Hurwitz, J. Nemirovsky, K. G. Makris, and M. Segev, "Experimental generation of arbitrarily shaped diffractionless superoscillatory optical beams.," Opt. Express **21**, 13425–35 (2013).
13. R. Remez and A. Arie, "Super-narrow frequency conversion," Optica **2**, 472–475 (2015).
14. G. Yuan, E. T. F. Rogers, T. Roy, L. Du, Z. Shen, and N. I. Zheludev, "Plasmonic Super-oscillations and Sub-Diffraction Focusing," in *CLEO: QELS_Fundamental Science* (Optical Society of America, 2014).
15. M. Dubois, E. Bossy, S. Enoch, S. Guenneau, G. Lerosey, and P. Sebbah, "Time-Driven Superoscillations with Negative Refraction," Phys. Rev. Lett. **114**, 13902 (2015).
16. Y. Eliezer and A. Bahabad, "Super-transmission: the delivery of superoscillations through the absorbing resonance of a dielectric medium," Opt. Express **22**, 31212–31226 (2014).
17. J. Verbeeck, H. Tian, and P. Schattschneider, "Production and application of electron vortex beams.," Nature **467**, 301–304 (2010).
18. N. Voloch-Bloch, Y. Lereah, Y. Lilach, A. Gover, and A. Arie, "Generation of electron Airy beams.," Nature **494**, 331–5 (2013).
19. V. Grillo, G. Carlo Gazzadi, E. Karimi, E. Mafakheri, R. W. Boyd, and S. Frabboni, "Highly efficient electron vortex beams generated by nanofabricated phase holograms," Appl. Phys. Lett. **104**, 43109 (2014).
20. C. Ophus, J. Ciston, J. Pierce, T. R. Harvey, J. Chess, B. J. McMorran, C. Czarnik, H. H. Rose, and P. Ercius, "Efficient linear phase contrast in scanning transmission electron microscopy with matched illumination and detector interferometry," Nat. Commun. **7**, 10719 (2016).
21. R. Shiloh, Y. Lereah, Y. Lilach, and A. Arie, "Sculpturing the electron wave function using nanoscale phase masks," Ultramicroscopy **144**, 26–31 (2014).
22. M. P. Cagigal, J. E. Oti, V. F. Canales, and P. J. Valle, "Analytical design of superresolving phase filters," Opt. Commun. **241**, 249–253 (2004).
23. D. B. Williams and C. B. Carter, *Transmission Electron Microscopy*, 2nd ed. (Springer US, 2009).
24. Z. S. Hegedus and V. Sarafis, "Superresolving filters in confocally scanned imaging systems," J. Opt. Soc. Am. A **3**, 1892 (1986).
25. B. J. McMorran, A. Agrawal, I. M. Anderson, A. A. Herzing, H. J. Lezec, J. J. McClelland, and J. Unguris, "Electron Vortex Beams with High Quanta of Orbital Angular Momentum," Science (80-. ). **331**, 192–195 (2011).
26. W. H. Lee, "Binary computer-generated holograms.," Appl. Opt. **18**, 3661–3669 (1979).
27. M. Uchida and A. Tonomura, "Generation of electron beams carrying orbital angular momentum," Nature **464**, 737–739 (2010).
28. R. E. Dunin-Borkowski and J. M. Cowley, "Simulations for imaging with atomic focusers," Acta Crystallogr. Sect. A Found. Crystallogr. **55**, 119–126 (1999).
29. J. M. Cowley, J. C. H. Spence, and V. V. Smirnov, "The enhancement of electron microscope resolution by use of atomic focusers," Ultramicroscopy **68**, 135–148 (1997).
30. H. Sawada, N. Shimura, F. Hosokawa, N. Shibata, and Y. Ikuhara, "Resolving 45-pm-separated Si–Si atomic columns with an aberration-corrected STEM," Microscopy **64**, 213–217 (2015).
31. B. K. Singh, H. Nagar, Y. Roichman, and A. Arie, "Particle manipulation beyond the diffraction limit using structured super-oscillating light beams," arXiv:1609.08858 (2016).
32. S. Lloyd, M. Babiker, and J. Yuan, "Quantized Orbital Angular Momentum Transfer and Magnetic Dichroism in the Interaction of Electron Vortices with Matter," Phys. Rev. Lett. **108**, 74802 (2012).

33. J. Verbeeck, H. Tian, and G. Van Tendeloo, "How to Manipulate Nanoparticles with an Electron Beam?," Adv. Mater. **25**, 1114–1117 (2013).

34. G. Yuan, S. Vezzoli, C. Altuzarra, E. T. F. Rogers, C. Couteau, C. Soci, and N. I. Zheludev, "Quantum super-oscillation of a single photon," arXiv **1510.03658**, (2015).

35. G. Yuan, E. T. F. Rogers, T. Roy, G. Adamo, Z. Shen, and N. I. Zheludev, "Planar super-oscillatory lens for sub-diffraction optical needles at violet wavelengths.," Sci. Rep. **4**, 6333 (2014).

36. K. G. Makris and D. Psaltis, "Superoscillatory diffraction-free beams," Opt. Lett. **36**, 4335–4337 (2011).

37. A. M. H. Wong and G. V Eleftheriades, "An Optical Super-Microscope for Far-field, Real-time Imaging Beyond the Diffraction Limit.," Sci. Rep. **3**, 1715 (2013).

38. E. T. F. Rogers and N. I. Zheludev, "Optical super-oscillations: sub-wavelength light focusing and super-resolution imaging," J. Opt. **15**, 94008 (2013).

39. D. Slepian and H. Pollak, "Prolate spheroidal wave functions, Fourier analysis and uncertainty—I," Bell Syst. Tech. J. **40**, 43–64 (1961).

40. P. Ben Dixon, D. J. Starling, A. N. Jordan, and J. C. Howell, "Ultrasensitive beam deflection measurement via interferometric weak value amplification," Phys. Rev. Lett. **102**, 173601 (2009).

41. Y. Aharonov, D. Z. Albert, and L. Vaidman, "How the result of a measurement of a component of the spin of a spin- 1/2 particle can turn out to be 100," Phys. Rev. Lett. **60**, 1351–1354 (1988).

42. T. Juffmann, A. Milic, M. Müllneritsch, P. Asenbaum, A. Tsukernik, J. Tüxen, M. Mayor, O. Cheshnovsky, and M. Arndt, "Real-time single-molecule imaging of quantum interference," Nat. Nanotechnol. **7**, 297–300 (2012).

**Supplemental Material for: Super-oscillating Electron Wave Functions with Sub-diffraction Spots**

Roei Remez[*†1], Yuval Tsur[†1], Peng-Han Lu[2], Amir H. Tavabi[2], Rafal E. Dunin-Borkowski[2] and Ady Arie[1]

[1] School of Electrical Engineering, Fleischman Faculty of Engineering, Tel-Aviv University, Israel
[2] Ernst Ruska-Centre for Microscopy and Spectroscopy with Electrons and Peter Grünberg Institute, Forschungszentrum Jülich, Jülich, Germany

[†] These authors contributed equally to this work
[*] Corresponding author: roei.remez@gmail.com

**Low convergence angle experiment**

The super-oscillating experiment was repeated for a small convergence angle (Fig. S1). We include this result, since factors such as inelastic scattering, spatial incoherence and chromatic aberration are weaker with respect to the spot size. Therefore, a better match with theory is apparent. This experiment was performed using low angle diffraction mode in a Tecnai F-20 FEG-TEM, where $\lambda = 2.5$ pm and the convergence semi-angle of the beam was $\alpha = 0.05$ mrad.

**Discussion: On the feasibility of a $10\lambda$ super-oscillation**

As suggested by Fig. 3a, a $10\lambda$ (=20 pm) radius electron hot-spot can be created in today's state of the art electron microscopes. Here, we discuss how spatial or temporal incoherence, as well as overall stability and signal to noise ratio, practically limit super-oscillation size using today's technology, which allows for the correction of coherent aberrations (astigmatism, spherical aberration etc.) within the aperture $\alpha = 32.4$ mrad [1].

**Spatial Incoherence.** The field emission gun tip is de-magnified when imaged onto the specimen plane to form a small probe. The size of the image of the gun determines the amount of smear expected for the measured wave function. In a state-of-the art STEM, this smear is about 7 pm [1]. As simulated in Fig. S2, the $10\lambda$ radius hot-spot described in Fig. 3a can indeed be realized experimentally using today's technology.

We stress that, even when the source is spatially incoherent, the wavefront of each emitted electron contains the super-oscillation (which can be arbitrarily small). However, the reduction in contrast caused by spatial incoherence would hamper the observation of a small super-oscillation in a large ensemble of electrons.

**Temporal Incoherence.** The energy uncertainty of the electron results in chromatic aberration. We split the discussion to two effects. First, the focal length of the lens changes with the wavelength of the electron, causing different electron wave functions to be slightly defocused compared to the designed wavelength. We performed simulations for a chromatic aberration coefficient Cc of 1.35 mm and an energy spread of $\Delta E = 0.5$ eV, which is determined as a sum of squares of the electron gun energy spread of 0.4 eV [1], a typical instability of the high-voltage supply of 0.2 eV and a typical instability of the condenser system, equivalent to 0.2 eV (with $\alpha = 25$ mrad, corresponding to Fig. 3a). As Fig. S3 shows, current chromatic aberrations cause a negligible smear of the hot-spot. Second, even if Cc is zero, the scale of the probe plane changes by an amount $\Delta\lambda/\lambda$, which is around $10^{-6} \ll 1$ and therefore gives a negligible smearing effect. We note that temporal incoherence can be reduced by using an electron monochromator [2], which transfer the problem to a signal to noise ratio issue (which can be compensated by using a large exposure time, assuming that the electronics and mechanics are stable).

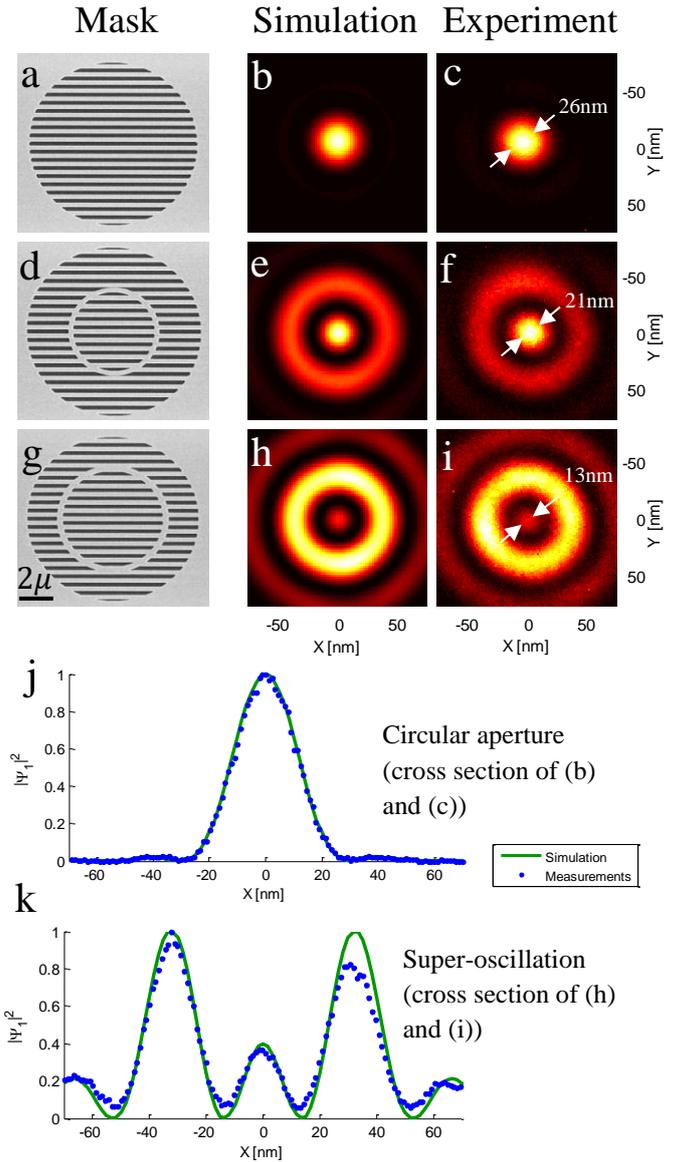

**Figure S1.** Binary amplitude masks (a,d,g), simulations (b,e,h) and experimental intensity patterns at the focal plane (c,f,i) for a circular aperture, a super-oscillation with $r_\pi = 0.52 r_{max}$ and a super-oscillation with $r_\pi = 0.65 r_{max}$. (j) and (k) are cross sections for a circular aperture and a super-oscillation with $r_\pi = 0.65 r_{max}$, respectively, all for $\alpha = 0.05$ mrad.

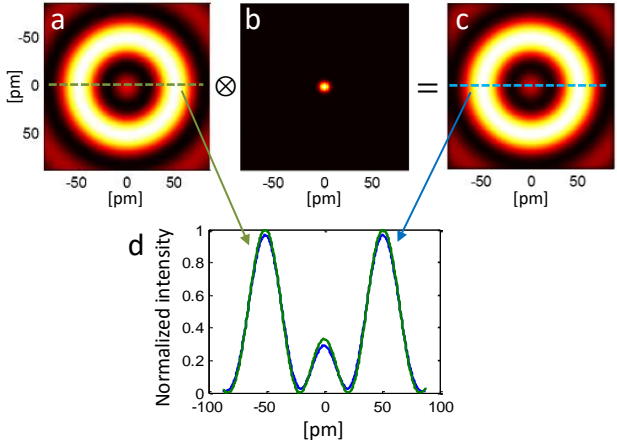

**Figure S2** – The effect of spatial incoherence on a $10\lambda$ super oscillating hot-spot for $\alpha = 25$ mrad. (a) The super oscillating probe assuming perfect coherence. (b) Gun shape after demagnification, having a FWHM of 7 pm. (c) The convolution of (a) and (b). (d) – Cross sections of (a) and (c).

**Inelastic Scattering.** Some of the contrast degradation apparent in Fig. 2 is attributed to inelastic scattering from the parts of the mechanically supporting 100 nm SiN, which were not milled. Removing this support (for example, see the mask in ref. [3]) would lead to the exclusion of any inelastic scattering, thus enabling a smaller spot.

**Overall Stability and Signal-to-Noise Ratio.** The intensity of the hot-spot in Fig. 3a is only three times lower than that of the peripheral ring. Therefore, Figs. 2 and 3a are not different in terms of signal-to-noise ratio. When trying to decrease the hot-spot size towards the de Broglie wavelength, significant limitations stem from both the low relative intensity of the hot-spot (as shown in Fig. S4) and the mechanical stability of the microscope column.

**Experimental details**

For the experimental realization of this concept, 200 nm SiN membranes were e-beam coated with a 150 nm Au layer. The Au layer and 100nm SiN were milled using a Ga focused ion beam machine (Raith IonLine). The masks were mounted in the C2 aperture plane of a probe-corrected electron microscope (FEI Titan 80-300 STEM [4]) operated in STEM mode. The probe was focused onto the specimen plane and then imaged on a post-column energy filter camera (Gatan Tridiem 865 ER). While the microscope had an aberration corrector for the pre-specimen, probe forming lenses, there was no such corrector for the post-specimen, imaging lenses. The post-specimen (imaging) lenses of the microscope had a spherical aberration of 1.2 mm.

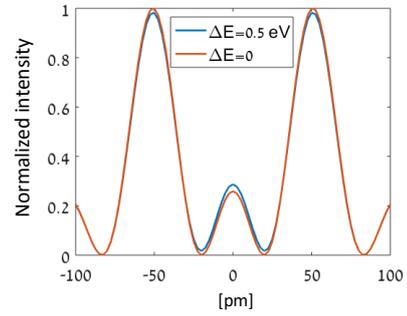

**Figure S3.** The effect of chromatic aberration on a $10\lambda$ super oscillating hot-spot. Parameters: $\alpha = 25$ mrad, $C_c = 1.35$ mm, $\Delta E = 0.5$ eV, E=300 KeV. Incoherent summation was carried out over an ensemble of electrons having a Gaussian energy distribution with $\sigma_E = \Delta E / 2\sqrt{2\ln(2)} = 0.21 eV$.

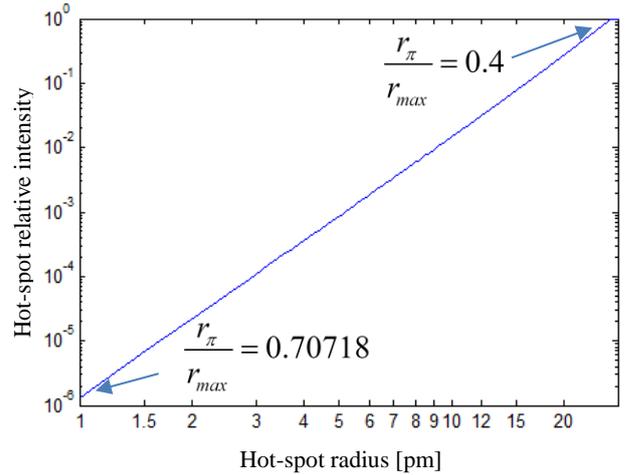

**Figure S4** – Hot-spot peak intensity divided by side-lobe peak intensity, shown as a function of hot-spot radius. The calculations were performed for a convergence semi-angle of 25 mrad and $\lambda = 2$ pm. Each point in this plot corresponds to a different value of $r_\pi / r_{max}$

**References**

1. H. Sawada, N. Shimura, F. Hosokawa, N. Shibata, and Y. Ikuhara, "Resolving 45-pm-separated Si–Si atomic columns with an aberration-corrected STEM," Microscopy **64**, 213–217 (2015).

2. Tiemeijer, P. C., M. Bischoff, B. Freitag, and C. Kisielowski. "Using a monochromator to improve the resolution in TEM to below 0.5 Å. Part I: creating highly coherent monochromated illumination." Ultramicroscopy **114,** 72-81 (2012).

3. J. Verbeeck, H. Tian, and P. Schattschneider, "Production and application of electron vortex beams.," Nature **467**, 301–304 (2010).

4. Heggen M, Luysberg M, and Tillmann K (2016). FEI Titan 80-300 STEM. Journal of large-scale research facilities 2 (2016) A42. http://dx.doi.org/10.17815/jlsrf-2-67